\documentclass[11pt,a4paper]{article}
\usepackage{jheppub}
\usepackage{amsmath}
\usepackage[most]{tcolorbox}
\usepackage{dsfont}
\usepackage{ulem}
\usepackage{natbib}
\usepackage{xcolor}
\usepackage[hang,flushmargin]{footmisc}
\usepackage{tikz-cd}
\usepackage{enumitem}
\usepackage{amsfonts,amssymb, amscd,amsmath,latexsym,amsbsy,bm}
\usepackage{stmaryrd}
\usepackage{todonotes}

\makeatletter\renewcommand{\@biblabel}[1]{#1.}\makeatother

\newtcolorbox{empheqboxed}{colback=gray!20, 
 colframe=white,
 width=\textwidth,
 sharpish corners,
 top=0mm, 
 bottom=0pt
}

\title{Gamma function solutions to the star-triangle equation}

\author{Ege Eren$^a$, Ilmar Gahramanov$^{a,b,c}$, Shahriyar Jafarzade$^{c,d}$ and Gonenc Mogol$^{e,f}$}
\affiliation{$^a$ Department of Physics, Bogazici University,\\ 34342 Bebek, Istanbul, Turkey\\[-0.5cm]

$^b$ Department of Mathematics, Khazar University, \\ Mehseti St. 41, AZ1096, Baku, Azerbaijan\\[-0.5cm]

$^{c}$ Institute of Radiation Problems, Azerbaijan National Academy of Sciences, \\ B.Vahabzade St. 9, AZ1143, Baku, Azerbaijan\\[-0.5cm]

$^{d}$ Institute of Physics, Jan Kochanowski University,\\ ul. Swietokrzyska 15, 25-406, Kielce, Poland \\[-0.5cm]

$^{e}$ Department of Physics and Astronomy, Stony Brook University,\\ Stony Brook, NY 11794, U.S.A. \\[-0.5cm]

$^{f}$ Department of Physics and Astronomy, Heidelberg University,\\ Heidelberg 69120, Germany \\

}

\emailAdd{ege.eren@boun.edu.tr}
\emailAdd{ilmar.gahramanov@boun.edu.tr}
\emailAdd{shahriyar.jzade@gmail.com}
\emailAdd{gonenc.mogol@stonybrook.edu}

\abstract{In the paper, we clarify some relations between solutions to the star-triangle equation via the gauge/YBE correspondence. We consider two solutions to the star-triangle relation in terms of Euler's gamma function. We derive these solutions from the reduction of certain basic and hyperbolic hypergeometric integral identities. These identities can be interpreted as equality of the supersymmetric partition functions of a specific three-dimensional $\mathcal N=2$ supersymmetric dual theories. 
}


\begin{document}
\maketitle
\flushbottom

\section{Introduction and conclusion}

There is a remarkable connection \cite{Spiridonov:2010em,Yamazaki:2012cp,Yamazaki:2013nra},  called gauge/YBE correspondence, between supersymmetric gauge theories and integrable lattice models of statistical mechanics. This correspondence is used to be quite useful in the construction of the new integrable lattice spin models \cite{Yamazaki:2013nra,Gahramanov:2015cva,Kels:2015bda, Yamazaki:2015voa,Gahramanov:2016ilb,Gahramanov:2017idz,Jafarzade:2017fsc,Kels:2017vbc, Yagi:2015lha}. Curiously, almost all known solutions to the star-triangle relation make an appearance in the context of this correspondence and the subject is under active investigation, see, e.g. \cite{Sarkissian:2018ppc,Kels:2018xge,Kels:2018xnf}. The purpose of this paper, by using the gauge/YBE correspondence is to interpret some integrable lattice spin models with Boltzmann weights in terms of Euler's gamma function from the supersymmetric gauge theory point of view. However we will not discuss details of the gauge/YBE correspondence here, more details can be found in the original works mentioned above and in the review papers \cite{Gahramanov:2017ysd,Yamazaki:2018xbx}. 

The sufficient condition for the integrability of the Ising-type lattice spin model (edge-interacting models) is the following star-triangle relation (the Yang-Baxter equation) for the Boltzmann weights \cite{Baxter:1982zz,Baxter:1987eq,Baxter:1997tn}
\begin{align} \nonumber
& \int S(\sigma) \overline{W}_{\beta}(\sigma,\sigma_j) W_{\gamma}(\sigma_k,\sigma) \overline{W}_{\alpha}(\sigma_i,\sigma) d \sigma \\ \label{STR1}
& \qquad \qquad \qquad \qquad \qquad = {\mathcal R}({\alpha,\beta,\gamma}) W_{\beta}(\sigma_k,\sigma_i) \overline{W}_{\gamma}(\sigma_i,\sigma_j) W_{\alpha}(\sigma_k,\sigma_j) ,  \end{align}
\begin{align} \nonumber
&  \int S(\sigma) \overline{W}_{\beta}(\sigma_j,\sigma) W_{\gamma}(\sigma,\sigma_k) \overline{W}_{\alpha}(\sigma,\sigma_i) d \sigma \\ \label{STR2}
& \qquad \qquad \qquad \qquad \qquad = {\mathcal R}(\alpha,\beta,\gamma) W_{\beta}(\sigma_i,\sigma_k) \overline{W}_{\gamma}(\sigma_j,\sigma_i) W_{\alpha}(\sigma_j,\sigma_k),
\end{align}
where $\sigma, \sigma_i \in \mathbb R$ stand for the spin variables\footnote{In discrete spin case, one needs to replace integration by summation.}, $\mathcal{W}_{\alpha}(\sigma_k,\sigma_j)$ and $\overline{\mathcal{W}}_{\alpha}(\sigma_k,\sigma_j)$   are two different kind of the Boltzmann weights (horizontal and vertical) with spectral parameter $\alpha$. $\mathcal{S}(\sigma)$ is the  single-spin self-interaction  term and $\mathcal{R}(\alpha,\beta,\gamma)$ is the spin-independent weight which can often be eliminated by some normalization of the Boltzmann weights. 

In this paper we will re-derive two solutions to the star-triangle relation presented in \cite{Kels:2013ola,Kels:2015bda} and in \cite{Bazhanov:2016ajm,Kels:2018xge}. The Boltzmann weights of both models are given in terms of Euler's gamma function. It turns out that the solutions presented in \cite{Kels:2013ola,Kels:2015bda} and \cite{Bazhanov:2016ajm,Kels:2018xge} can be obtained from higher-level hypergeometric solutions. From the supersymmetric gauge theory side, we explicitly work out the reduction procedure of three-dimensional $\mathcal N=2$ supersymmetric sphere partition function and superconformal index to the two-dimensional $\mathcal N=(2,2)$ sphere partition function for the supersymmetric dual theories. From the mathematical point of view, we make several reductions of the basic and hyperbolic hypergeometric integral identities to the ordinary hypergeometric integral identities.

 The first model has the following vertical and horizontal Boltzmann weights \cite{Bazhanov:2016ajm,Kels:2018xge}
\begin{equation}
  \label{sol1.1}
  \overline{W}_{\alpha}(x,z) = \Gamma\left( \alpha \pm i x \pm i z  \right) \qquad W_{\alpha}(x,z) = \frac{ \Gamma\left(- \alpha + i x \pm i z\right)  }{\Gamma\left(+ \alpha + i x \pm i z\right)} \;,
\end{equation}
the following self-interaction term and the spin-independent coefficient
\begin{equation} \label{sol1.2}
  S(z) = \frac{1}{\Gamma(\pm2 i z)} \quad \text{and} \quad {\mathcal R}(\alpha, \beta, \gamma) =  \frac{4\pi\Gamma(2\alpha)\, \Gamma(2\beta)}{\Gamma(2\gamma)}  \;.
\end{equation}
We use the notation that $\pm$ signs in special functions indicate a product of functions, for instance, $\Gamma(x\pm y)=\Gamma(x+y) \Gamma(x-y)$. In the context of the gauge/YBE correspondence, this solution can be obtained from the equality of two-dimensional vortex partition functions for a certain dual theories. 

The second integrable lattice spin model which we discuss here has the following Boltzmann weight \cite{Kels:2013ola,Kels:2015bda}
\begin{align}\label{sol2.1}
 {W}_{\alpha}(\sigma_i|m_i,\sigma_j|m_j)=\frac{\Gamma(\frac{1+\alpha}{2})}{\Gamma(\frac{1-\alpha}{2})}\,
\frac{\Gamma(\frac{1-\alpha}{2}\pm\frac{i(\sigma_i+\sigma_j)-(m_i+m_j)}{2})\,\Gamma(\frac{1-\alpha}{2}\pm\frac{i(\sigma_i-\sigma_j)-(m_i-m_j)}{2})}
{\Gamma(\frac{1+\alpha}{2}\pm\frac{i(\sigma_i+\sigma_j)+(m_i+m_j)}{2})\,\Gamma(\frac{1+\alpha}{2}\pm\frac{i(\sigma_i-\sigma_j)+(m_i-m_j)}{2})},
  \end{align} 
and self-interaction term 
  \begin{align} \label{sol2.2}
      \mathcal{S}(\sigma,m)=\frac{\sigma^2+m^2}{2\pi} \;.
  \end{align}
In this model the spin-independent parameter is equal to one due to the suitable normalization of the Boltzmann weights. The Boltzmann weights (\ref{sol2.1})-(\ref{sol2.2}) solve the star-triangle relation of the following form
\begin{align}\nonumber
\sum_{m\in \mathbb{Z}} \int d\sigma\, \mathcal{S}( \sigma|m)&\mathcal{W}_{ \eta - \alpha}( \sigma _i|m_i , \sigma |m)\mathcal{W}_{ \eta - \beta}( \sigma _j|m_j, \sigma |m) \mathcal{W}_{ \eta - \gamma}( \sigma _k|m_k, \sigma |m) \\ \label{STR}
 = & \mathcal{W}_{  \alpha}( \sigma _j|m_j, \sigma _k|m_k)W_{ \beta}( \sigma _i|m_i, \sigma _k|m_k) \mathcal{W}_{  \gamma}( \sigma _j|m_j, \sigma _i|m_i) \;,
\end{align}
where the crossing parameter $\eta=\alpha+\beta+\gamma$. One can obtain the relation (\ref{STR}) from (\ref{STR1}) by restricting the Boltzmann weights, see, e.g. \cite{auyang201630, perk2006yang}. Further we will see that there are several ways to obtain the solution (\ref{sol2.1})-(\ref{sol2.2}).

The main goal of the paper is to show connections between different solutions to the star-triangle relation explicitly. 

It would be interesting to consider also the reductions of the ${\mathbb RP}^2 \times S^1$ supersymmetric partition function \cite{Tanaka:2015pwa, Mori:2015urc,Tanaka:2014oda,Bozkurt:2018xno}, where shrinking $S^1$ gives a matrix integral also in terms of gamma functions.

It is also possible to generalize the reductions discussed here to other dualities, see, e.g. \cite{Jafarzade:2018yei} and obtain new or known integral identities in terms of the gamma function, expressing the equivalence of dual partition functions. There are several solutions to the Yang-Baxter equation in terms of gamma functions \cite{au1999large,Bazhanov:2007vg,Derkachov:2016dhc,Derkachov:2016ucn,Derkachov:2019ynh}, it would be interesting to consider other models in this context. 

\begin{figure}[h]
\begin{center}
    \includegraphics[width=0.75\linewidth]{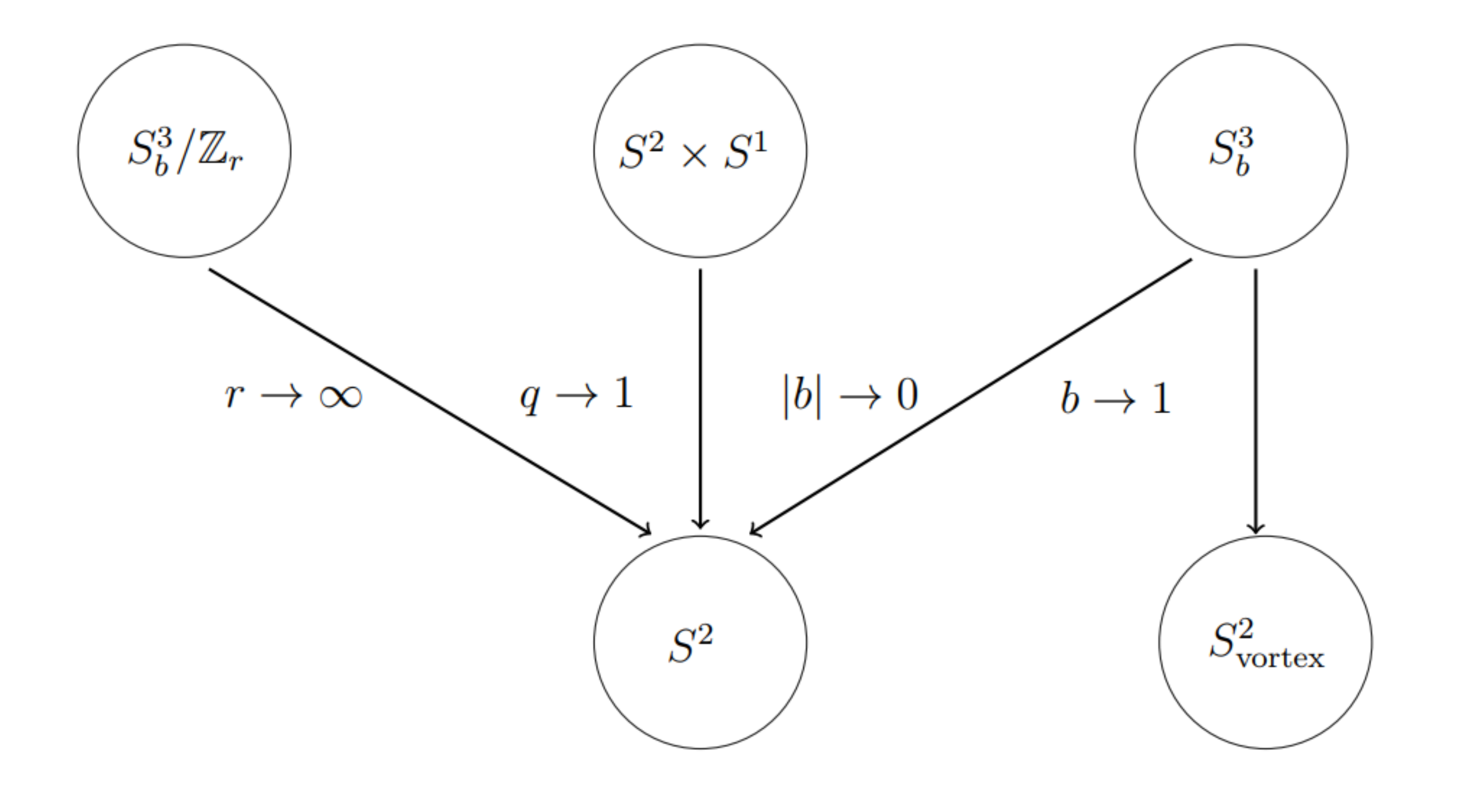}
\caption{Structure of the paper}
\end{center}
\end{figure}

The diagram in Fig. 1 demonstrates the plan of the paper pictorially. The rest of the paper is organized as follows:

\begin{itemize}
   \item A brief review of the necessary information on supersymmetric partition functions and $\mathcal N=2$ supersymmetric duality are contained in Sections 2 and 3.  
\item The relations between different solutions to the star-triangle relation and the main results are presented in Sec. 4.

    \item In Sec. 4.1 we present the reduction of $S_b^3$ partition function to the two-dimensional vortex partition function and obtain the solution (\ref{sol1.1})-(\ref{sol1.2}) to the star-triangle relation. In \cite{Kels:2013ola}, Kels presented another limit $b \rightarrow 0$ of the three-dimensional squashed sphere partition function, which gives the two-dimensional sphere partition function. 
    
    \item The reduction of $S^2 \times S^1$ partition function (three-dimensional superconformal index) to $S^2$ partition function ($S^1$ shrinks to zero size) and the solution (\ref{sol2.1})-(\ref{sol2.2}) to the star-triangle equation are discussed in Sec. 4.2. We make the same limiting procedure discussed in the paper \cite{Kels:2015bda} using slightly different notations, see also \cite{Gahramanov:2017ysd}. 
    
    \item In Sec. 4.3 we consider the $r \rightarrow \infty$ limit of the $S^3/\mathbb{Z}_r$ partition function. This reduction was discussed in \cite{Benini:2011nc}, here we worked out the limit in detail on the level of special functions and obtain the solution (\ref{sol2.1})-(\ref{sol2.2}).
    
\end{itemize}

\section{Supersymmetric partition functions}

Let us briefly summarize the basic ingredients which one needs to know about three-dimensional theories with four supercharges ($\mathcal N=2$ theories). The theory contains the vector multiplets consisting of a gauge field, a complex Dirac fermion, a real scalar field, and an auxiliary scalar field belong to the adjoint representation of gauge group, and chiral multiplets consisting of a complex scalar field, a complex Dirac fermion, and a complex auxiliary belong to a suitable representation of gauge group and flavor group.  The supersymmetry algebra contains the $SO(2)$ $R$-symmetry which rotates supercharges. In the context of gauge/YBE correspondence, the R-charge plays the role of spectral parameter.

The partition function of three-dimensional $\mathcal N=2$ supersymmetric gauge theory on a certain compact manifold can be computed exactly due to the supersymmetric localization technique \cite{Pestun:2007rz}. By using the localization one ends with the following matrix integral\footnote{Here we consider the partition function as a Coulomb branch integral, in general one can use different localization locus, see, for instance, the Higgs branch localization \cite{Fujitsuka:2013fga,Benini:2013yva,Nieri:2015yia}}
\begin{align}\label{SUSY-PF}
\mathcal{Z}=\frac{1}{|W|}\sum \oint \prod_{\text{Cartan}} [dz] \;  \mathcal{Z}_{\text{vector}} \; \mathcal{Z}_{ \text{chiral}}\;,
\end{align}
Here $\mathcal{Z}_{\text{vector}}$ and $\mathcal{Z}_{ \text{chiral}}$ stand for the contribution of the vector and chiral multiplets, respectively. The integral is performed over the Cartan subgroup of the gauge group. $|W|$ represents the order of the Weyl group of thr gauge group. In our examples, we will consider theories without the Chern–Simons and Fayet-Iliopoulos term\footnote{Actually, it seems that in some cases Fayet-Iliopoulos term plays an important role the integrability, see \cite{Yamazaki:2016wnu}.}, therefore we skip contributions of these terms to (\ref{SUSY-PF}). The meaning of summation and integration will be explained in the next sections for each special case.

\subsection{Three-dimensional \texorpdfstring{$\mathcal N=2$}{N=2} superconformal index}

Three-dimensional $\mathcal N=2$ superconformal index\footnote{In the path-integral formulation superconformal index is a partition function of a theory defined on the background $S^2 \times S^1$.} was studied in \cite{Imamura:2011su,Kim:2009wb,Krattenthaler:2011da}. The one-loop contribution of chiral multiplets to the index is given by
\begin{equation}
Z_{\rm chiral}= \prod_{j} \prod_{\rho_j } \prod_{\phi_j} \frac{(q^{1-\frac{\Delta_j}{2}+\frac{|\rho_j(m)+ \phi_j(n)|}{2}-\rho_j(z)-\phi_j(\Phi)}; q)_{\infty}}{(q^{\frac{\Delta_j}{2}+\frac{|\rho_j(m)+\phi_j(n)|}{2}+\rho_j(z)+\phi_j(\Phi)}; q)_{\infty}} \;,
\end{equation}
where $j$ labels chiral multiplets,  $\rho_j,\phi_j$, are the weights of the representation of the   gauge  and  flavor groups, respectively and $\Delta_j$ is the Weyl weight of $j$'th chiral multiplet. Here we use the usual q-Pochhammer symbol 
\begin{equation}
(z;q)_{\infty} = \prod_{j=0}^{\infty} (1-z q^j) \;.
\end{equation}
The one-loop contribution of the vector multiplet combined with the Vandermonde determinant, is given also in terms of q-Pochammer symbols
\begin{align} \nonumber
Z_{\rm vector} = \prod_{\alpha \in R_+} (q^{\alpha(z)+|\alpha(m)|/2};q)_\infty (q^{-\alpha(z)+|\alpha(m)|/2};q)_\infty \;,
\end{align}
where the product is over the positive roots $\alpha$ of the gauge group $G$.

For the case of superconformal index  the summation in (\ref{SUSY-PF}) is over the GNO quantized fluxes \cite{Goddard:1976qe} (monopole charges) $m_i=\frac{1}{2\pi} \int_{S^2}F_i$.

\subsection{Squashed three-sphere partition function}

$\mathcal N=2$ supersymmetric partition function on the squashed three-sphere $S_b^3$ was studied in \cite{Benini:2011nc, Imamura:2012rq, Nieri:2015yia}. The one-loop contribution of chiral multiplets to the partition function can be expressed as
\begin{equation}
Z_{\rm chiral}=
\prod_{j} \prod_{\rho_j} \prod_{\phi_j} \gamma^{(2)}\left(\frac{
Q}{2} \Delta_j + \rho_j(z)+\phi_j({\Phi}); \omega_1, \omega_2\right) \;,
\end{equation}
where $j$ labels chiral multiplets,  $\rho_j,\phi_j$, are the weights of the representation of the   gauge  and  flavor groups, respectively and $\Delta_j$ is the Weyl weight of $j$'th chiral multiplet. Here  $Q=b+\frac{1}{b}$ with the squashing parameter $b^2=\omega_2/\omega_1$. The function  $\gamma^{(2)}(u, \omega_1,\omega_2)$ is the hyperbolic gamma function\footnote{We should mention that when the squasing parameter is real, the infinite product representation (\ref{gammatwo}) is not valid, and one needs to use the integral representation, see \cite{Spiridonov:2010em}.} defined as follows \cite{BultThesis}
\begin{equation} \label{gammatwo}
\gamma^{(2)}(u;\omega_1, \omega_2)=e^{-\pi i B_{2,2}(u;\mathbf{\omega_1,\omega_2})/2} \frac{(e^{2 \pi i u/\omega_1}\tilde{q};\tilde{q})}{(e^{2 \pi i u/\omega_1};q)} \quad \text{with} \quad q=e^{2 \pi i \omega_1/\omega_2}, \quad \tilde{q}=e^{-2 \pi i \omega_2/\omega_1} \; ,
\end{equation}
where  $B_{2,2}(u;\omega_1, \omega_2)$ is the second order Bernoulli polynomial,
\begin{equation} B_{2,2}(u;\omega_1,\omega_2) =
\frac{u^2}{\omega_1\omega_2} - \frac{u}{\omega_1} -
\frac{u}{\omega_2} + \frac{\omega_1}{6\omega_2} +
\frac{\omega_2}{6\omega_1} + \frac 12.
\end{equation}
The one-loop contribution of the vector multiplet combined with the Vandermonde determinant, is given by
\begin{align} \nonumber
Z_{\rm vector} = \prod_{\alpha \in R_+} \frac{1}{\gamma^{(2)}\left(\alpha(z); \omega_1, \omega_2\right)
\gamma^{(2)}\left(-\alpha(z); \omega_1, \omega_2 \right)} \; ,
\end{align}
where the product is over the positive roots $\alpha$ of the gauge group $G$.

Note that the squashed partition function has a form of (\ref{SUSY-PF}) without summation.

\subsection{Orbifold partition function}

Supersymmetric partition function on $S_b^3/Z_r$, called orbifold partition function, was studied in \cite{Benini:2011nc, Imamura:2012rq, Nieri:2015yia, Gahramanov:2016ilb}. The one-loop contribution of chiral multiplets to the partition function is given by\footnote{
In supersymmetry literature, usually one uses the other special function, namely, the contribution of the chiral multiplet has a form

\begin{equation}
Z_{\rm chiral}=
\prod_{j}\prod_{\rho_j}\prod_{\phi_j}
\hat s_{b,-\rho_j( m)-\phi_j( n)} 
\left(i\frac{Q}{2} (1-\Delta_j)-\rho_j(z)-\phi_j( \Phi)\right)~,
\end{equation}
where the function  $\hat s_{b,-m}$ is the improved double sine function \cite{Nieri:2015yia}
\begin{equation} \label{shat}
    \hat s_{b,-m}(x)=\sigma(m)\,e^{\frac{\pi i}{2}B_\varphi(x,m,\omega_1,\omega_2)}\prod_{j=0}^{r-1}\frac{(e^{2\pi(x+i \omega_2\llbracket m\rrbracket)/(\omega_2r)}\,(e^{\pi i(\omega_1+\omega_2)/(\omega_2r)})^{2j+1};e^{2\pi i\omega_1/(\omega_2r)})_\infty}{(e^{2\pi(x- i\omega_1\llbracket m\rrbracket)/(\omega_1r)}\,(e^{-\pi i(\omega_1+\omega_2)/(\omega_1r)})^{2j+1};e^{-2\pi i\omega_2/(\omega_1r)})_\infty}\,,
\end{equation}
where $\llbracket  m\rrbracket _r\in\{0,1,\ldots ,r-1\}$ denotes $m \mbox{ modulus } r$, $\sigma(m)=e^{\frac{ i \pi}{2r}(\llbracket m\rrbracket(r-\llbracket m\rrbracket)-(r-1)m^2)}$ and $B_\varphi$ is a particular combination of multiple Bernoulli polynomials given by
\begin{equation}
\label{bphidef}
B_\varphi(z,m,\omega_1,\omega_2):=B_{2,2}(i z+\omega_1\llbracket m \rrbracket+\eta,r\omega_1,2\eta)+B_{2,2}(i z+\omega_2(r-\llbracket m\rrbracket)+\eta,r\omega_2,2\eta)\;.
\end{equation}
The one-loop contribution of the vector multiplet is given by
\begin{align} \nonumber
Z_{\rm vector} = \prod_{\alpha \in R_{+}}\frac{1}{\hat s_{b, \alpha(m)} 
\left( i \frac{Q}{2}+ \alpha(z)\right) \hat s_{b, -\alpha(m)} 
\left( i \frac{Q}{2}- \alpha(z)\right)} \;.
\end{align}
}
\begin{equation}
Z_{\rm chiral}=
\prod_{j}\prod_{\rho_j}\prod_{\phi_j}
 \Gamma_h\left(\frac{Q}{2} (1-\Delta_j)+\rho_j(z)+\phi_j( \Phi), \rho_j( m)+\phi_j( n) \right)~,
\end{equation}
where $j$ labels chiral multiplets,  $\rho_j,\phi_j$, are the weights of the representation of the   gauge  and  flavor groups, respectively and $\Delta_j$ is the Weyl weight of $j$'th chiral multiplet. Here $Q=b+\frac{1}{b}$ with the squashing parameter $b^2=\omega_2/\omega_1$. The function  $\Gamma_h$ is a version of the improved double sine function \cite{Gahramanov:2016ilb}, which can be written as a product of hyperbolic gamma functions
\begin{align*}
    \Gamma_h(z,m;\omega_1,\omega_2)=e^{\phi_h(m)}\gamma_h(z,m;\omega_1,\omega_2) \;,
\end{align*}
where 
\begin{align}
    \phi_h(m)= -\frac{\pi i}{6r}\big(2m^3-3m^2r+mr^2\big) \;,
\end{align}
and
\begin{align}
    \gamma_h(z,m;\omega_1,\omega_2)=\gamma^{(2)}\Big(-iz-i\omega_2\big(r-m\big);-i\omega_2r,-i\omega_1-i\omega_2\Big)\\\nonumber
         \qquad\times\gamma^{(2)}\Big(-iz-i\omega_1m;-i\omega_1r,-i\omega_1-i\omega_2\Big) \;.
\end{align}

The one-loop contribution of the vector multiplet combined with the Vandermonde determinant, is given by
\begin{align} \nonumber
Z_{\rm vector} = \prod_{\alpha \in R_{+}}\frac{1}{\Gamma_h \left(\alpha(z), \alpha(m)\right) \Gamma 
\left(\alpha(z), -\alpha(m)\right)} \; ,
\end{align}
where the product is over the positive roots $\alpha$ of the gauge group $G$.

In the orbifold partition function case the summation in the formula (\ref{SUSY-PF}) is over holonomies $m_i=\frac{r}{2\pi}\int_{C}A_{\mu}dx^{\mu}$, where the integration countour is over non-trivial cycle on $S^3_b/Z_r$ and $A_\mu$ is the gauge field.

\subsection{Two-dimensional sphere partition function}

The $\mathcal N=(2,2)$ supersymmetric partition function on $S^2$ was obtained in 
\cite{Benini:2012ui,Doroud:2012xw}. The one-loop contribution of chiral multiplets to the partition function is given by
\begin{align}
    \mathcal{Z}_{\text{chiral}}=\prod_{j}\prod_{\rho_j}\prod_{\phi_j} \frac{\Gamma(\frac{\Delta_j}{2}-i\rho_j(z) - \phi_j(\Phi)-\frac{1}{2}\rho_j (m))}{\Gamma(1-\frac{\Delta_j}{2}+i\rho_{j}( z)+\phi_j(\Phi)+\frac{1}{2}\rho_{j} (m))} \; ,
\end{align}
where $j$ labels chiral multiplets,  $\rho_j,\phi_j$, are the weights of the representation of the   gauge  and  flavor groups, respectively and $\Delta_j$ is the Weyl weight of $j$'th chiral multiplet.  Here $\Gamma(z)$ function is the usual Euler's gamma function.

The one-loop contribution of the vector multiplet for theory with non-abelian gauge group combined with the Vandermonde determinant, is given by
\begin{align}
\mathcal{Z}_{\text{vector}}=\prod_{\alpha \in R_{+}}(-1)^{m}\Big(\frac{\alpha(m)^2}{4}+\alpha(z)^2\Big) \;,
\end{align}
where the product is over the positive roots $\alpha$ of the gauge group $G$.

In the two-dimensional sphere partition function case, the integration contour in formula (\ref{SUSY-PF}) is defined along the real lines  and summation is taken over the magnetic fluxes.

\section{Supersymmetric duality}

For our purposes we consider the following three-dimensional $\mathcal N=2$ supersymmetric duality \cite{Gahramanov:2016wxi}:

The first theory is the SQCD with $SU(2)$ gauge group and with $SU(6)$ flavor group, chiral multiplets are in the fundamental representation of the gauge group and the flavor group, a vector multiplet is in the adjoint representation of the gauge group.

The second theory, i.e. the dual one, has no gauge degrees of freedom, fifteen chiral multiplets of the theory are in the totally antisymmetric tensor representation of the flavor group.

One of the tools for checking supersymmetric dualities is to compute the partition function, which is expected to be the same for dual theories. 

The equality of the superconformal indices for dual theories can be expressed in terms of the basic hypergeometric sum and the integral identity \cite{Gahramanov:2016wxi}
\begin{align}\label{qSeib} \nonumber
     \sum_{m=-\infty}^{\infty}  \oint \prod_{j=1}^6 \frac{(q^{1+(m+n_j)/2}/g_jz,q^{1+(n_j-m)/2} z/g_j ;q)_\infty}{(q^{(m+n_j)/2}g_jz,q^{(n_j-m)/2} g_j/z ;q)_\infty} \frac{(1-q^{m} z^2)(1-q^{m} z^{-2})}{q^{m} z^{6m}} \frac{dz}{2\pi i z} \\ =\frac {2}{ \prod_{j=1}^6 g_j^{n_j}}  \prod_{1 \leq j<k  \leq 6} \frac {(q^{1+(n_j+n_k)/2}/g_j g_k;q)_\infty}{(q^{(n_j+n_k)/2}g_j g_k;q)_\infty} \; , 
\end{align}
with the conditions $\prod_{j=1}^6 g_i = q;$ and $\sum_{j=1}^6 n_j =  0$. 

The integral identity for the squashed sphere partition functions can be expressed in terms of the following hyperbolic hypergeometric functions \cite{BultThesis}
\begin{equation}\label{eq:gt_integral}
  \int_{-\infty}^\infty
  \frac{\prod_{j=1}^6\gamma^{(2)}(g_k\pm i z;\omega_1,\omega_2) }
  {\gamma^{(2)}(\pm 2 i z;\mathbf{\omega_1,\omega_2}) } d z
  =2\sqrt{\omega_1\omega_2}
  \prod_{1\leq j<k\leq 6}\gamma^{(2)}(g_j+g_k;\mathbf{\omega_1,\omega_2}) \;,
\end{equation}
where the parameters $g_{j}$'s obey the balancing condition $
  \sum_{j=1}^{6} g_{j} = \omega_{1} + \omega_{2}$.

The integral identity for the duality at the level of orbifold partition functions has the following form \cite{Gahramanov:2016ilb}
\begin{align} \nonumber
    \frac{1}{2r\sqrt{-\omega_1\omega_2}}\sum_{m=0}^{\lfloor r/2 \rfloor}\epsilon (m) & \int _{-\infty}^{\infty} dz \frac{\prod_{j=1}^6 \Gamma_h(g_j\pm z,n_j\pm m; \omega_1,\omega_2 )}{\Gamma_h(\pm 2z,\pm 2m; \omega_1,\omega_2 )} \\
    & \qquad   =\prod_{1\leq j<k\leq 6} \Gamma_h(g_j\pm g_k,n_j\pm n_k; \omega_1,\omega_2 ) \; ,
\end{align}
with the balancing condition $\sum_{j}g_j=\omega_1+\omega_2$ and $\epsilon(0)=1$ and $\epsilon(m)=2$ for $m>0$.

All these integral identities can be written in the form of the star-triangle relation \cite{Spiridonov:2010em,Gahramanov:2015cva,Gahramanov:2016ilb, Kels:2015bda}. The solutions (\ref{sol1.1})-(\ref{sol1.2}) and (\ref{sol2.1})-(\ref{sol2.2}) to the star-triangle equation arise from the reductions of the above identities.

\section{Solutions to the star-triangle equation}

\subsection{Reduction of \texorpdfstring{$S^3_b$}{S3} partition function}

In order to obtain the solution (\ref{sol1.1})-(\ref{sol1.2})  to the star-triangle equation from Spiridonov's generalization of the Faddeev-Volkov model, we will consider the reduction of the three-dimensional supersymmetric squashed sphere partition function to the identity for the two-dimensional supersymmetric vortex partition function. This reduction procedure was studied in \cite{Spiridonov2014}.

Now let us reduce the identity (\ref{eq:gt_integral}) to the hyprgeometric level. Using the balancing condition for the identity (\ref{eq:gt_integral}), we rewrite one of the fugacities (in this case $g_6$) as  $g_6= 2\eta - \sum_{j=1}^{5} g_{j}$  and use the reflection identity for the hyperbolic gamma function
\begin{equation}\label{Ref-Id}
\gamma^{(2)}(z;\omega_1,\omega_2) \; \gamma^{(2)}(\omega_1+\omega_2-z;\omega_1,\omega_2) = 1,
\end{equation}
to bring terms containing $g_6$ to the denominator on both sides of the identity \eqref{eq:gt_integral}. Then we apply the following reduction formula of the hyperbolic gamma function
\begin{equation}\label{omega-limit}
    \gamma^{(2)}(z;\omega_2)\underset{\omega_2\rightarrow \infty} {=} \Big(\frac{\omega_2}{2\pi \omega_1}\Big)^{\frac{1}{2}-\frac{z}{\omega_1}}\frac{\Gamma(z/\omega_1)}{\sqrt{2\pi}}
\end{equation}
to the integral identity  (\ref{eq:gt_integral}) and obtain\footnote{Actually, here we have used the asymptotic behaviour of $\gamma^{(2)}(z, \omega_1,\omega_2)$ function, 
\begin{align}\label{As-beh}
\lim_{z \rightarrow \infty}
e^{\frac{\pi \textup{i}}{2} B_{2,2}(z;\omega_1,\omega_2)} \gamma^{(2)}(z; \omega_1,\omega_2)
& =  1, \text{ \ \ for } \text{arg }\omega_1 < \text{arg } z < \text{arg }\omega_2 + \pi,  \\\nonumber
\lim_{z \rightarrow \infty}e^{-\frac{\pi \textup{i}}{2} B_{2,2}(z;\omega_1,\omega_2)} \gamma^{(2)}(z;\omega_1,\omega_2)
& =  1, \text{  \ \ for } \text{arg } \omega_1 - \pi < \text{arg } z < \text{arg }\omega_2.
\end{align}
and canceled a factor of \((\omega_{2} \big/ 2\pi \omega_{1} )^{3}\), which appears on both sides of the equation (\ref{Gamma-int}).}
\begin{align} \nonumber
   \int   \frac{ \prod_{k=1}^{5} \Gamma\left( \frac{g_{k} \pm i z}{\omega_{1}}  \right)  } { \Gamma(\pm 2i z/\omega_{1}  ) \cdot \Gamma\left( 1/ {\omega_{1}} \cdot \left( \pm i z + \sum_{k=1}^{5} g_{k}   \right) \right)  }\,  d z \\ \label{Gamma-int}
 \qquad   =
  4\pi\omega_1\frac{ \prod\limits_{1 \leq j < k \leq 5} \Gamma((g_{j}+g_{k})/\omega_{1})  }{\prod_{j=1}^{5} \Gamma\left( 1/\omega_{1} \cdot \left( -g_{j} +\sum_{k=1}^{5} g_{k}   \right)  \right)} \; .
\end{align}

Now we introduce a new fugacity $g_6$ via the following equation (a new balancing condition)
\begin{equation}
  \label{eq:balancing_cond_gamma}
  \sum_{j=1}^{6} g_{j} = 0 \; .
\end{equation}
Then the integral identity (\ref{Gamma-int}) can be rewritten as
\begin{equation}
  \label{eq:gamma_int}
  \int  \frac{ \prod_{k=1}^{5} \Gamma\left( \frac{g_{k} \pm i z}{\omega_{1}}  \right)  } { \Gamma(\pm 2i z/\omega_{1}  ) \cdot \Gamma\left( \frac{- g_{6} \pm i z } {\omega_{1}} \right)  }\, d z =  4\pi\omega_1
  \frac{ \prod\limits_{1 \leq j < k \leq 5} \Gamma((g_{j}+g_{k})/\omega_{1})  }{\prod_{j=1}^{5} \Gamma\left( -  \frac{g_{j}+g_{6}}{\omega_{1}}  \right)} \; .
\end{equation}
We also introduce the following parameters
\begin{equation}
  \label{eq:gamma_param}
  g_{1,2} = + \alpha \pm i \sigma_i \;, \quad g_{3,4} = + \beta \pm i \sigma_j \;, \quad g_{5,6} = -\gamma \pm i \sigma_k \;, \quad z = \sigma
\end{equation}
and the Boltzmann weights
\begin{equation}
  \label{eq:boltzman_weights_gamma}
  \overline{W}_{\alpha}(\sigma_i,\sigma_j) = \Gamma\left( \frac{\alpha \pm i \sigma_i \pm i \sigma_j}{\omega_{1}}  \right) \qquad W_{\alpha}(\sigma_i,\sigma_j) = \frac{ \Gamma\left(\frac{- \alpha + i \sigma_i \pm i \sigma_j}{\omega_{1}}\right)  }{\Gamma\left(\frac{+ \alpha + i \sigma_i \pm i \sigma_j}{\omega_{1}}\right)} 
\end{equation}
\begin{equation} \label{eq:self_norm_gamma}
  S(\sigma) = \frac{1}{2\pi \Gamma(\pm2 i \sigma)} \quad \text{and} \quad R(\alpha, \beta, \gamma) =  \frac{2\omega_{1}\Gamma(2\alpha)\, \Gamma(2\beta)}{\Gamma(2\gamma)} \; . 
\end{equation}
Noting that the parameterization \eqref{eq:gamma_param} and the equation \eqref{eq:balancing_cond_gamma} implies \(\alpha + \beta - \gamma = 0 \). Then, by choosing $\omega_1 = 1$, we can write the integral identity \eqref{eq:gamma_int} as
\begin{equation}\label{eq:st_gamma_fst}
  \int S(\sigma) \overline{W}_{\beta}(\sigma,\sigma_j) W_{\gamma}(\sigma_k,\sigma) \overline{W}_{\alpha}(\sigma_i,\sigma) d \sigma = R({\alpha,\beta,\gamma}) W_{\beta}(\sigma_k,\sigma_i) \overline{W}_{\gamma}(\sigma_i,\sigma_j) W_{\alpha}(\sigma_k,\sigma_j) \; .
\end{equation}
The equation \eqref{eq:st_gamma_fst} is one of the two non-symmetric star-triangle equations \eqref{sol1.1} and \eqref{sol1.2}. 

We will show that the Boltzmann weights defined as above also satisfy the second star-triangle equation. First, by using the reflection identity (\ref{Ref-Id})  we rewrite the integral identity \eqref{eq:gt_integral}  as follows:

\begin{multline}
  \int \frac{ \prod_{j=1}^{4} \gamma^{(2)}(g_{j} \pm i z) \,\gamma^{(2)}(g_{5}+i z)\, \gamma^{(2)}(2\eta - \sum_{j=1}^{5} g_{j} + i z)  }{ \gamma^{(2)}(\pm 2 i z) \, \gamma^{(2)}( 2\eta - g_{5} + i z  ) \gamma^{(2)}(\sum_{j=1}^{5} g_{j} + i z) } d z
  = \\ 
  2\sqrt{\omega_1\omega_2} \, \frac
  { \left(\prod_{j=2}^{5} \gamma^{(2)}(g_{1}+ g_{j}) \right) \, \gamma^{(2)}(g_{1}+2\eta - \sum_{j=1}^{5} g_{j}) \, \gamma^{(2)}(g_{2}+g_{3})}
  { \gamma^{(2)}(2\eta - g_{2}-g_{5}) \, \gamma^{(2)}(-g_{2}+\sum_{j=1}^{5} g_{j}) \,\gamma^{(2)}(2\eta-g_{4}-g_{5})  }   \\\times 
 \frac{ \gamma^{(2)}(g_{2}+g_{4})\,\gamma^{(2)}(g_{3}+g_{4}) \, \gamma^{(2)}(g_{3}+g_{5}) \, \gamma^{(2)}(g_{3}+ 2\eta - \sum_{j=1}^{5} g_{j}) }
  {\gamma^{(2)}(-g_{4}+\sum_{j=1}^{5} g_{j}) \, \gamma^{(2)}(-g_{5} + \sum_{j=1}^{5} g_{j})  } \; ,
\end{multline}

where we have defined a shorthand notation $\gamma^{(2)}(z)=\gamma^{(2)}(z;\omega_1,\omega_2)$ and $2\eta=\omega_1+\omega_2$. Since the above integral is valid for all values of \(\omega_{1}\) and \(\omega_{2}\), we can choose \(\omega_{2} = \omega_{1} \cdot n\) for some $n\in \mathbb{N}$. Notice that with this choice, we get $(2\eta/ \omega_{1} = (1 + n) \in \mathbb{N})$. Then we use the asymptotic formula (\ref{omega-limit}) for the $\gamma^{(2)}(z,\omega_1,\omega_2)$ function when \(n \to \infty\) and obtain the following identity
\begin{multline} \label{4.12}
    \int \frac{ \prod_{j=1}^{4} \Gamma(\frac{g_{j} \pm i z}{\omega_1}) \, \Gamma(\frac{g_{5}+i z}{\omega_1})\, \Gamma(\frac{g_{6} + i z}{\omega_1})  }{ \Gamma(\frac{\pm 2 i z}{\omega_1}) \, \Gamma(\frac{ - g_{5} + i z}{\omega_1}  ) \Gamma(\frac{-g_{6} + i z}{\omega_1}) } d z
  =4\pi \omega_1 \frac
  {
    \left(\prod_{j=2}^{6} \Gamma(\frac{g_{1}+ g_{j}}{\omega_1}) \right) \, \Gamma(\frac{g_{2}+g_{3}}{\omega_1})\, \Gamma(\frac{ g_{2}+g_{4}}{\omega_1})}
  { \Gamma(\frac{ - g_{2}-g_{5}}{\omega_1}) \, \Gamma(\frac{ -g_{2}-g_{6}}{\omega_1}) \,\Gamma(\frac{-g_{4}-g_{5}}{\omega_1})  }   \\ \times
  \frac{\Gamma(\frac{ g_{3}+g_{4}}{\omega_1}) \, \Gamma(\frac{ g_{3}+g_{5}}{\omega_1}) \, \Gamma( \frac{g_{3}+ g_{6}}{\omega_1}) }
  {\Gamma(\frac{ -g_{4}-g_{6}}{\omega_1}) \, \Gamma(\frac{ -g_{5}-g_{6}}{\omega_1})  } \; ,
\end{multline}
where we used the property of the Euler gamma function
\begin{equation}
  \Gamma\left(\frac{2\eta}{\omega_{1}} - \frac{\Sigma + i z}{\omega_{1}}\right) = \left({2\eta}\big/{\omega_{1}}\right)! \cdot \Gamma\left( -\frac{\Sigma+ i z}{\omega_{1}}  \right) \;.
\end{equation}

Again, we make a choice $\omega_1 = 1$ and finally obtain the second star-triangle equation from the integral identity (\ref{4.12}) by using the parameterization \eqref{eq:gamma_param} and the definition of Boltzmann weights \eqref{eq:boltzman_weights_gamma}

\begin{equation}\label{eq:st_gamma_snd}
  \int S(\sigma) \overline{W}_{\beta}(\sigma_j,\sigma) W_{\gamma}(\sigma,\sigma_k) \overline{W}_{\alpha}(\sigma,\sigma_i) d \sigma = R(\alpha,\beta,\gamma) W_{\beta}(\sigma_i,\sigma_k) \overline{W}_{\gamma}(\sigma_j,\sigma_i) W_{\alpha}(\sigma_j,\sigma_k). 
\end{equation}

\subsection{Reduction of \texorpdfstring{$S^2\times S^1$}{S2 x S1} partition function}

From the supersymmetric gauge theory of view, shrinking the circle\footnote{The shrinking procedure on the level of supersymmetric partition function was studied first for four-dimensional theories \cite{Dolan:2011rp,Gadde:2011ia,Imamura:2011uw} (see also \cite{Gahramanov:gka,Gahramanov:2015tta,Gahramanov:2013xsa}).} $S^1$ to zero gives rise to a two–dimensional supersymmetric theory with the same amount of supercharges on $S^2$ \cite{Benini:2011nc, Aharony:2017adm, Benini:2012ui}. Computationally, we use the following limit of the q-Pochammer symbol
\begin{equation}
        \lim_{q \rightarrow 1} \frac {(q^\alpha;q)_\infty}{(q^\beta;q)_\infty} (1-q)^{\alpha-\beta}= \frac{ \Gamma (\beta)}{ \Gamma (\alpha)}.
\end{equation}
By applying this formula to the integral identity (\ref{qSeib}), one obtains the following expression
\begin{align}\nonumber
  \sum_{m\in\mathbb{Z}}\int \frac{dz}{2\pi} \frac{\Gamma( m\pm 2iz+1)}{\Gamma(m\pm2iz)} \prod_{j=1}^{6} \frac{\Gamma(\frac{m+n_j}{2}+g_j+ iz)}{\Gamma(1+\frac{ m+n_j}{2}-g_j- iz)}\frac{\Gamma(\frac{-m+n_j}{2}+g_j- iz)}{\Gamma(1+\frac{ -m+n_j}{2}-g_j+ iz)}\\ \label{Kelssolution2}
   =\prod_{1\leq j<k\leq6} \frac{\Gamma(g_j+g_k+\frac{n_j+n_k}{2})}{\Gamma(1-g_j-g_k-\frac{n_j+n_k}{2})}.
\end{align}
Plugging the following substitutions for the fugacities 
\begin{align} \nonumber
  g_{1,4}=\frac{\alpha}{2}\pm \frac{i\sigma_i}{2};\qquad g_{2,5}=\frac\beta 2\pm\frac{i\sigma_j}{2}; \qquad g_{3,6}=\frac\gamma 2\pm\frac{i\sigma_k}{2}; \qquad iz=\frac{i\sigma_0}{2} 
\end{align}
and following constraints on monopole charges
\begin{align} 
   n_1=-n_4, \quad n_2=-n_5, \quad n_3=-n_6,
\end{align}
we obtain the solution to the star-triangle relation with discrete and continuous spin variables
\begin{align}\nonumber
\mathcal{W}_{\alpha}(\sigma_j,n_j|\sigma_k,n_k)=\frac{\Gamma(\frac{n_k+n_j}{2}+\frac{\eta-\alpha}{2}+\frac{i\sigma_j+ i\sigma_k}{2})}{\Gamma(1+\frac{ n_k+n_j}{2}-\frac{\eta-\alpha}{2}-\frac{i\sigma_j+ i\sigma_k}{2})}\frac{\Gamma(\frac{-n_k+n_j}{2}+\frac{\eta-\alpha}{2}+\frac{i\sigma_j-i\sigma_k}{2})}{\Gamma(1+\frac{ -n_k+n_j}{2}-\frac{\eta-\alpha}{2}-\frac{i\sigma_j-i\sigma_k}{2})} \\
\times
\frac{\Gamma(\frac{n_k-n_j}{2}+\frac{\eta-\alpha}{2}-\frac{i\sigma_j- i\sigma_k}{2})}{\Gamma(1+\frac{ n_k-n_j}{2}-\frac{\eta-\alpha}{2}+\frac{i\sigma_j-i\sigma_k}{2})}\frac{\Gamma(\frac{-n_k-n_j}{2}+\frac{\eta-\alpha}{2}-\frac{i\sigma_j+i\sigma_k}{2})}{\Gamma(1+\frac{ -n_k-n_j}{2}-\frac{\eta-\alpha}{2}+\frac{i\sigma_j+i\sigma_k}{2})} 
\end{align}
 \begin{equation}
       \mathcal{S}(\sigma,m)=\frac{1}{2\pi}\frac{\Gamma(m\pm i\sigma+1)}{\Gamma(m\pm i\sigma)},
\end{equation}
\begin{align}\label{R} 
\mathcal{R}(\alpha, \beta, \gamma)=\frac{\Gamma(\alpha)}{\Gamma(\eta-\alpha)} \frac{\Gamma(\beta)}{\Gamma(\eta-\beta)}\frac{\Gamma(\gamma)}{\Gamma(\eta-\gamma)}.
\end{align}
This is exactly the solution found in \cite{Kels:2013ola}.

\subsection{Reduction of \texorpdfstring{$S_b^3/Z_r$}{S3/Zr} partition function}

In this section we consider the Euler Gamma function limit by taking $b=1$ and  $r\rightarrow \infty$ from the solution corresponding to $S_b^3/Z_r$ partition function. In order to do it, we use the limit from  $S^3/Z_r$ to  $S^2$ partition function.

For $\omega_1=\omega_2=\omega$ performing the asymptotic formula for the $\gamma^{(2)}(z)$ function, one finds the following formula
\begin{align}
    \gamma_h(z,m;\omega_1,\omega_2) \underset{{ r\rightarrow \infty}}{=} \big(\frac{r}{2\pi}\big)^{-\frac{z}{\omega}+1}\frac{\Gamma(\frac{z}{2\omega}+\frac{m}{2})}{\Gamma(1-\frac{z}{2\omega}+\frac{m}{2})} \;.
\end{align}
By normalizing the fugacities as $z \rightarrow \frac{z}{2i\omega}$ and $t_i \rightarrow \frac{g_i}{2\omega}$, we obtain the following integral identity
\begin{align} \nonumber
   \sum_{m=0}^{\infty}\epsilon (m) & \int _{-\infty}^{\infty}   \frac{dz}{2\pi } \prod_{j=1}^6\frac{\Gamma(g_j+ iz+\frac{n_j+ m}{2})}{\Gamma(1-g_j- iz+\frac{n_j+ m}{2})}
    \frac{\Gamma(g_j- iz+\frac{n_j- m}{2})}{\Gamma(1-g_j+ iz+\frac{n_j- m}{2})}  \\  \label{intfromorbi}
   & \times \frac{\Gamma(1- 2iz+ m)}{\Gamma( 2iz+ m)} \frac{\Gamma(1+ 2iz- m)}{\Gamma(- 2iz- m)} 
   =   \prod_{1\leq j<k\leq 6}  \frac{\Gamma(g_j+ g_k+\frac{n_j+ n_k }{2})}{\Gamma(1-g_j- g_k +\frac{n_j+ n_k}{2})}
\end{align}
It is not difficult to show that this identity is equivalent to the integral identity (\ref{Kelssolution2}) (see Appendix A), and therefore the identity (\ref{intfromorbi}) gives the same solution to the star-triangle equation, i.e. one can obtain the solution (\ref{sol2.1})-(\ref{sol2.2}).

From the special function point of view, an alternative method  leading to the reduction of $S^3/\mathbb{Z}_r$ partition function to the $S^2$ partition function is described in \cite{Benini:2011nc}.

\section*{Acknowledgement}
 SJ is thankful to ICTP and Simons Center for Geometry and Physics for the warm hospitality during the program ``Exactly Solvable Models of Quantum Field Theory and Statistical Mechanics'' where some parts of the work were done. The research on this project has received funding from the European Research Council (ERC) under the European Union's Horizon 2020 research and innovation program (QUASIFT grant agreement 677368) during the visit of Ilmar Gahramanov to the Institut des Hautes Etudes Scientifiques (IHES), where some of the research for this paper was performed. This work has supported by BAP Project (no. 2019-26), funded by Mimar Sinan Fine Arts University, Istanbul, Turkey.

\appendix

\section{Appendix}
Here we present the equality of integal identities (\ref{Kelssolution2}) and (\ref{intfromorbi}). Noting the definition of the $\epsilon(m)$ function,
\begin{equation}
\label{eq:A.1}
        \epsilon(m) = 
    \begin{cases} 
      1 & m = 0 \\
      2 & m > 0 \\
   \end{cases}
\end{equation}
left hand side of (\ref{intfromorbi}) can be rewritten as
\begin{multline}
    \label{eq:A.2}
    2\sum_{m=1}^{\infty}\int_{-\infty}^{\infty}\frac{dz}{2\pi}\frac{ \Gamma(1-i z+m)\Gamma(1+2iz-m)}{\Gamma(-2iz-m)\Gamma(2iz+m)}\prod_{j=1}^{6}\frac{\Gamma(g_j + iz + \frac{m_j+m}{2})\Gamma(g_j-iz+\frac{m_j-m}{2})}{\Gamma(1-g_j-iz+\frac{m_j+m}{2})\Gamma(1-g_j+iz+\frac{m_j-m}{2})} \\
    + \int_{-\infty}^{\infty}\frac{dz}{2\pi}\frac{ \Gamma(1-i z+m)\Gamma(1+2iz-m)}{\Gamma(-2iz-m)\Gamma(2iz+m)}\prod_{j=1}^{6}\frac{\Gamma(g_j + iz + \frac{m_j+m}{2})\Gamma(g_j-iz+\frac{m_j-m}{2})}{\Gamma(1-g_j-iz+\frac{m_j+m}{2})\Gamma(1-g_j+iz+\frac{m_j-m}{2})} \Bigg|_{m=0} \; .
\end{multline}
As can be observed, the left-most term in above equation remains unchanged if the change of variables $m \xrightarrow{} -m$ and $z \xrightarrow{} -z$ are made. We make use of this fact and express (\ref{eq:A.2}) as 
\begin{align}
    \sum_{m=1}^{\infty}\int_{-\infty}^{\infty}\frac{dz}{2\pi}\frac{ \Gamma(1-i z+m)\Gamma(1+2iz-m)}{\Gamma(-2iz-m)\Gamma(2iz+m)}\prod_{j=1}^{6}\frac{\Gamma(g_j + iz + \frac{m_j+m}{2})\Gamma(g_j-iz+\frac{m_j-m}{2})}{\Gamma(1-g_j-iz+\frac{m_j+m}{2})\Gamma(1-g_j+iz+\frac{m_j-m}{2})} \\ 
    +\sum_{m=1}^{\infty}\int_{-\infty}^{\infty}\frac{dz}{2\pi}\frac{ \Gamma(1-i z+m)\Gamma(1+2iz-m)}{\Gamma(-2iz-m)\Gamma(2iz+m)}\prod_{j=1}^{6}\frac{\Gamma(g_j + iz + \frac{m_j+m}{2})\Gamma(g_j-iz+\frac{m_j-m}{2})}{\Gamma(1-g_j-iz+\frac{m_j+m}{2})\Gamma(1-g_j+iz+\frac{m_j-m}{2})} \\
    + \int_{-\infty}^{\infty}\frac{dz}{2\pi}\frac{ \Gamma(1-i z+m)\Gamma(1+2iz-m)}{\Gamma(-2iz-m)\Gamma(2iz+m)}\prod_{j=1}^{6}\frac{\Gamma(g_j + iz + \frac{m_j+m}{2})\Gamma(g_j-iz+\frac{m_j-m}{2})}{\Gamma(1-g_j-iz+\frac{m_j+m}{2})\Gamma(1-g_j+iz+\frac{m_j-m}{2})} \Bigg|_{m=0} \\
    = \sum_{m=-\infty}^{\infty}\int_{-\infty}^{\infty}\frac{dz}{2\pi}\frac{ \Gamma(1-i z+m)\Gamma(1+2iz-m)}{\Gamma(-2iz-m)\Gamma(2iz+m)}\prod_{j=1}^{6}\frac{\Gamma(g_j + iz + \frac{m_j+m}{2})\Gamma(g_j-iz+\frac{m_j-m}{2})}{\Gamma(1-g_j-iz+\frac{m_j+m}{2})\Gamma(1-g_j+iz+\frac{m_j-m}{2})}
\end{align}
and complete the proof.

\bibliographystyle{utphys}
\bibliography{refYBE}



\end{document}